\documentclass[english]{article}
\usepackage[T1]{fontenc}
\usepackage[latin1]{inputenc}
\usepackage{babel}
\usepackage{graphics}

\makeatletter

\providecommand{\LyX}{L\kern-.1667em\lower.25em\hbox{Y}\kern-.125emX\@}

 \newcommand{\lyxaddress}[1]{
   \par {\raggedright #1 
   \vspace{1.4em}
   \noindent\par}
 }

\makeatother
\begin{document}

\title{\( \textrm{ } \)Dynamic Transitions in Pure Ising Magnets under
Pulsed and Oscillating Fields}

\author{Bikas K. Chakrabarti\( ^{1,2} \) and Arkajyoti Misra\( ^{2} \)}

\maketitle

\lyxaddress{\( ^{1} \)Saha Institute of Nuclear Physics, 1/AF Bidhan Nagar,
Kolkata 700064, India.}

\lyxaddress{\( ^{2} \)Theoretische Physik, Gerhard-Mercator-Universität Duisburg,
D-47048, Germany.}

\begin{abstract}
Response of pure Ising systems to time-dependent external magnetic
fields, like pulsed and oscillating fields, are discussed and compared
here. Because of the two time scales involved, namely the thermodynamic
relaxation time of the system and the pulse width or the time period
of the external field, dynamically broken symmetric phases appear
spontaneously when both become comparable. A particularly simple case
is that of an Ising ferromagnet below its static critical temperature,
when it is perturbed for a short duration by a pulsed magnetic field
competing with the existing order in the system. If the field strength
and duration is more than the threshold (dependent on the temperature),
the system, and consequently the magnetization, switches from one
minimum to the other of the static free energy. This magnetization
reversal transition here shows intriguing dynamic transition behaviour,
similar to those for oscillating fields. Monte Carlo studies for such
dynamic transitions are discussed and compared with the mean field
results for the same and the Monte Carlo results for the oscillating
field case. In particular, we discuss about the Monte Carlo results
for the fluctuations and their growth behaviour near this magnetization
reversal (dynamic) transition point.
\end{abstract}

\section{Introduction}

The response of pure ferromagnetic Ising systems under the action
of time-dependent external magnetic fields are being studied intensively
these days. A whole new class of dynamic phase transitions emerged
from the study of such driven spin systems under different time dependences
of the driving field. Originally it was seen that when a pure Ising
system is driven by an oscillating external field, it settles to a
dynamic steady state with spontaneously broken symmetry when the time
period of the external field becomes smaller compared to the system's
relaxation time. In this steady state, the period averaged magnetization
acquires a net non-zero value, while the period averaged magnetic
field remains zero \cite{ca99, lp90, ac95, srn98}. The universality
class of this dynamic transition has recently been established, from
extensive Monte Carlo studies, to be the same as that of the equilibrium
transition in the pure Ising system \cite{srn98, kwrn00}. Later,
efforts were made to investigate and analyse the response of such
systems under a pulsed magnetic field which, for a short duration,
either favours the existing magnetic order of the system (positive
pulse) or competes with it (negative pulse). The results of the positive
pulse case were analyzed by extending appropriately the finite size
scaling technique to this finite time-window case, and it did not
indicate any new phase transition or thermodynamic scale \cite{abc97}.
However, the negative pulse case indicated a new dynamic transition
and thermodynamic scales, as the appropriate pulse width and strength
just manages to help the system overcome the {}``double-well free-energy
barrier'' and the magnetization reversal transition occurs \cite{mc98, smc, mc00}.

In the next section we briefly introduce the model and the magnetization
reversal transition in this negative pulse case. We also briefly summarize
here a few analytic results obtained earlier in the mean field limit.
In section 3, we report Monte Carlo results for the fluctuations near
this magnetization reversal transition in two dimensional Ising systems.
In the concluding section, we analyse these observations and also
compare them with those obtained for the Ising model under oscillating
field.

\section{Model}

The model we consider here is the Ising model with nearest-neighbour
interaction under a time-dependent external magnetic field, described
by the Hamiltonian

\begin{equation}
\label{hamil}
H=-\frac{J}{2}\sum _{\left[ ij\right] }S_{i}S_{j}-h(t)\sum _{i}S_{i}.
\end{equation}
Here, \( J \) denotes the strength of the cooperative interaction
between nearest neighbour spins put under a time-dependent field \( h(t). \)
\( S_{i} \) and \( S_{j} \) denote the spins at lattice sites \( i \)
and \( j \) respectively and the nearest-neighbour spin pairs are
denoted by {[}...{]}. We consider such a system at temperatures \( T\textrm{ } \)
below its static critical temperature \( T_{c} \). The external field
is applied at \( t=t_{0} \) after the system is brought to equilibrium,
having an equilibrium magnetization \( m_{0} \) (\( T \)). The external
field \( h(t) \) has a time dependence given by \( h(t)=-h_{p} \)
for \( t_{0}\leq t\leq t_{0}+\Delta t \) and \( h(t)=0 \) outside
this range. Typical responses of the time-dependent magnetization
\( m(t) \) of the system under different \( h(t) \) are indicated
in Fig. 1. For appropriate combinations of \( h_{p} \) and \( \Delta t \),
at a fixed \( T \), a magnetization reversal transition occurs when
the magnetization of the system eventually switches to \( -m_{0} \).
It may noted that a magnetization reversal phenomenon trivially occurs
in the limit \( \Delta t\rightarrow \infty  \) for any non-vanishing
value of \( h_{p} \) at any temperature \( T<T_{c} \) . This is
however a limiting case of the transition, which is studied here for
any finite value of \( \Delta t \) . Also, in the case of magnetization
reversal transition, the magnetization reversal need not occur during
the presence of the external pulsed field. In fact, the time required
for the system to reach its final equilibrium state diverges as the
transition point is approached \cite{mc98}. 

A qualitative understanding for these diverging time and also length
scales at this magnetization reversal transition point can be easily
understood from a mean field study of this problem \cite{mc98, smc}.
Using mean field approximation for the dynamics, we can write the
equation of motion for the average magnetization \( m_{i}(=<S_{i}>) \)
as\begin{equation}
\label{mfeq1}
\frac{\mathrm{d}m_{i}}{\mathrm{d}t}=-m_{i}+\tanh \left( \frac{J\sum _{j}m_{j}+h(t)}{T}\right) ,
\end{equation}
 where \( <...> \) denotes the thermodynamic (ensemble) average.
This equation may be linearized near the magnetization reversal transition
point and for uniform magnetization one gets 

\begin{equation}
\label{solm(t)}
m(t)=m_{0}-\left( \frac{h_{p}}{\Delta T}-m_{0}\right) \left[ \exp \left\{ \frac{\Delta T}{T}\left( t-t_{0}\right) \right\} \right] 
\end{equation}
as a solution of Eq. (\ref{mfeq1}). Here \( \Delta T=T_{c}^{\mathrm{mf}}-T \),
where \( T_{c}^{\mathrm{mf}}\equiv J(q=0) \) is the critical temperature
in the mean field approximation and \( J(q) \) is the Fourier transform
of the interaction \( J \). From the solution (\ref{solm(t)}) the
value of \( m_{0} \) is always restricted to be less than \( h_{p}/\Delta T \)
in order to make the linear approximation valid. Hence during the
presence of the field, \( m(t) \) will decrease in its magnitude
from \( m(t_{0})\equiv m_{0} \) to \( m(t_{0}+\Delta t)\equiv m_{w} \)
at the time of withdrawal of the pulse. Since there is no fluctuation
here, the magnetization will relax back to its original value \( m_{0} \)
if \( m_{w} \) is positive or relax to a value \( -m_{0} \) if \( m_{w} \)
is negative. This can be easily seen by solving eq. (\ref{mfeq1})
for \( t>t_{0}+\Delta t \) where \( h(t)=0 \), using \( m_{w} \)
as the initial value of the magnetization. In this time region one
gets a Debye type relaxation of magnetization to its final equilibrium
value, with a relaxation time \cite{mc98}

\[
\tau \sim \left( \frac{1}{T_{c}^{\mathrm{mf}}-T}\right) \ln \left( \left| \frac{m_{0}}{m_{w}}\right| \right) .\]
 It diverges at the magnetization reversal transition point, where
\( m_{w} \) vanishes. The prefactor, giving the divergence of \( \tau  \)
at the static mean field transition temperature, is responsible for
the usual critical slowing down phenomenon at the static transition
point (\( h=0) \). The other factor gives the diverging time scale
of interest here, at any temperature below the static transition temperature,
where magnetization reversal occurs or \( m_{w} \) vanishes due to
the appropriate combination of the external field pulse strength and
duration. A similar solution for the Fourier transform of the linearized
version of the equation (\ref{mfeq1}), with an additional derivative
with respect to the field \( h \) for the susceptibility \( \chi (q) \),
gives \cite{smc, mc00} \[
\chi (q)\sim \exp \left( -q^{2}\xi ^{2}\right) ,\]
 where \[
\xi \sim \left[ \frac{1}{\left( T_{c}^{\mathrm{mf}}-T\right) }\ln \left( \frac{1}{\left| m_{w}\right| }\right) \right] ^{\frac{1}{2}}.\]
 Here also, the prefactor in \( \xi  \) gives the usual divergence
at \( T_{c}^{\mathrm{mf}} \) , while the other factor gives the divergence
of interest here at the magnetization reversal transition point. Such
divergences of \( \tau  \) and \( \xi  \) across the dynamic transition
phase boundary which gives the optimal combination of \( h_{p} \)
and \( \Delta t \) at any particular \( T \), are also seen \cite{mc00}
in the Monte Carlo simulations.

We believe, a similar transition and the associated length and time
scales are responsible in an essential way for the spontaneous dynamic
symmetry breaking transition in the Ising systems under oscillating
fields \cite{ca99, kwrn00}. The growing time scale induced by the
periodically changing field competing with the existing order in the
system at that point, locks the system dynamically in a broken symmetric
phase. In order to compare these scenario, we have studied here the
growing fluctuations in the order parameter \( m_{w} \) and the internal
energy \( E \), as one approaches the magnetization reversal phase
boundary in a Monte Carlo study. We find that the fluctuations in
both these quantities diverge very prominently at the magnetization
reversal phase boundary. We also indicate the behaviour of the derivatives
\( (\textrm{d}m_{w}/\textrm{d}h_{p}) \) and \( (\textrm{d}E/\textrm{d}T) \),
which show some peak across the same phase boundary. We compare our
results with those reported for the dynamic transition under oscillating
field \cite{a98, kwrn00}.

\section{Monte Carlo study}

We study here the behaviour of the fluctuations near the magnetization
reversal transition in a ferromagnetic Ising system on square lattice
of size \( L \), under negative or competing field pulses, using
Monte Carlo simulations with single spin flip Glauber dynamics. Working
at temperatures \( T \) below the static critical temperature \( T_{c} \)
(\( \simeq  \) 2.27 in units of \( J \) ), the system is brought
to the equilibrium state by evolving the system without any external
field, from an initial configuration (say all spins up). The equilibrium
relaxation time required for this system to evolve to its equilibrium
state with magnetization \( m_{0} \) is noted. The field pulse of
strength \( -h_{p} \) for a duration \( \Delta t \) (measured in
Monte Carlo steps or MCS), competing with this order, is applied at
\( t=t_{0} \) only after the system has reached its equilibrium.
The magnetization \( m(t) \) starts decreasing (from its equilibrium
value \( m_{0} \)) due to the effect of the competing field \( -h_{p} \)
during the time period \( \Delta t \). The average value of the magnetization
\( m_{w} \) at the time of the withdrawal of the pulse is noted.
The phase boundary of this dynamic transition gives the appropriate
combination of \( h_{p} \) and \( \Delta t \) that induces the magnetization
reversal by bringing \( m(t) \) from \( m(t_{0})=m_{0} \) to \( m(t_{0}+\Delta t)\equiv m_{w}=0 \).
Therefore across the phase boundary \( m_{w} \) changes its sign
\cite{mc98}. Obviously the phase boundary changes with \( T \).
We study the behaviour of various thermodynamic quantities across
these phase boundaries. The thermodynamic quantities are averaged
over different initial configurations of the system. We note the fluctuations
over these average quantities as well. In our simulations we have
considered three different system sizes, viz., \( L \) =256, 512
and 1024 for all our studies reported here. The typical number of
averages is 200 for \( L \) = 1024 and 1000 for \( L \) = 256. 

In Fig. 2 we show the growth of fluctuations \( \Delta m_{w}=L^{2}(<m_{w}^{2}>-<m_{w}>^{2}) \)
in \( m_{w} \) across the phase boundary at two different points,
both driven by the pulse strength \( h_{p} \). The insets here show
the corresponding variations of the pulse-withdrawal-time magnetization
\( m_{w} \), which crosses zero at the phase boundary. Figure 3 shows
the growth of fluctuations \( \Delta E_{\mathrm{tot}\textrm{ }}=L^{2}(<E_{\mathrm{tot}}^{2}>-<E_{\mathrm{tot}}>^{2}) \)
in the average total energy \( E_{\mathrm{tot}} \) per spin pair
(in units of \( J) \) across the phase boundary, again at two points,
both driven by \( T \). The insets here again show the corresponding
\( m_{w} \) variations. Both these fluctuations \( \Delta m_{w} \)
and \( \Delta E_{\mathrm{tot}} \) clearly diverge at the phase boundary,
in the large system size limit. Fig. 4 shows the variation of the
fluctuation \( \Delta E_{\mathrm{coop}} \) \( =L^{2}(<E_{\mathrm{coop}}^{2}>-<E_{\mathrm{coop}}>^{2}) \)
in the cooperative part \( E_{\mathrm{coop}} \) (excluding the field
contribution) of the total energy \( E_{\mathrm{tot}} \) for the
same parameter values as those of Fig. 3(a). Although the fluctuation
in \( E_{\mathrm{coop}} \) also grows in general around the transition
point, a sharp dip is observed there. This dip is identified to be
due to the sharp growth of \( E_{\mathrm{coop}} \) at the transition
point, as shown in the inset.  

In Fig. 5, we show the behaviour of the derivatives \( \textrm{d}m_{w}/\textrm{d}h_{p} \)
(in (a)) and \( \textrm{d}E_{\mathrm{tot}}/\textrm{d}T \) (in (b))
across the phase boundary at two different points. Unlike in the case
of the dynamic transition under oscillating fields \cite{a98}, we
do not observe any divergence in these quantities at the transition
point. This is because the curves are independent of system size and
hence the peak values do not grow with \( L \). However, the peaks
become sharper at lower temperatures (higher values of \( h_{p} \))
as the system approaches the tricritical point occurring on the phase
boundary \cite{mc00}.

\section{Concluding remarks}

We have studied here the growth of fluctuations and their system size
dependence across the dynamic magnetization reversal phase boundary.
This transition can be driven by the field pulse strength \( h_{p} \)
or its duration \( \Delta t \) or even by varying \( T \) in the
presence of a fixed field pulse but always keeping it below \( T_{c} \).
`Critical slowing down' and the divergence of length scales etc at
this dynamic phase boundary have already been demonstrated earlier
\cite{mc00}, using both mean field approximation and the Monte Carlo
simulations in two and three dimensions. Here we have shown the Monte
Carlo results for the growth of fluctuations, both in the field-withdrawal-time
magnetization \( m_{w} \) and in the total energy \( E_{\mathrm{tot}\textrm{ }} \),
across the dynamic phase transition boundary in two dimension. The
growth of these fluctuations with the system size \( L \) clearly
indicates a divergence at the phase boundary. It may also be noted
that although we have shown the fluctuations for one specific driving
parameter, e.g. \( \Delta m_{w} \) with \( h_{p} \) and \( \Delta E_{\mathrm{tot}} \)
with \( T \), similar fluctuations in both the quantities are seen
for other driving parameters as well. All these clearly indicate a
genuine dynamic phase transition at the magnetization reversal point
of an Ising system under a competing field pulse.

Acknowledgement: BKC is grateful to  K. D. Usadel for his kind hospitality
in the Theoretical Physics Group, University of Duisburg, where a
part of the work was done and the manuscript was finalized. A part
of the work has been supported by the Deutsche Forschungsgemeinschaft
through SFB 491 and Graduiertenkolleg 277.

\textbf{\LARGE Figure Captions:}{\LARGE \par}
\vspace{2cm}

Fig. 1 Typical time variation of the response magnetizations \( m(t) \)
for two different field pulses \( h(t) \) with different strengths
\( h_{p} \), but same width \( \Delta t \) and at same temperature
\( T \), are shown. The quantities of interest to characterize the
response magnetization are also indicated. 

Fig. 2 Growth of fluctuations \( \Delta m_{w} \) near the transition,
driven by \( h_{p} \), are shown for two different points on the
phase boundary. (a) for \( T \) = 1.0 and \( \Delta t \) = 12 MCS,
and (b) for \( T \) = 1.50 and \( \Delta t \) = 2 MCS. The insets
show the corresponding \( m_{w} \) variations; \( m_{w} \) = 0 at
the transition point. 

Fig. 3 Growth of fluctuations \( \Delta E_{\mathrm{tot}} \) near
the transition, driven by \( T \) , are shown for two different points
on the phase boundary. (a) for \( h_{p} \) = 1.75 and \( \Delta t \)
= 10 MCS, and (b) for \( h_{p} \) = 1.25 and \( \Delta t \) = 10
MCS. The insets show the corresponding \( m_{w} \) variations; \( m_{w} \)
= 0 at the transition point.

Fig. 4 Growth of fluctuations \( \Delta E_{\mathrm{coop}} \) near
the transition point at \( h_{p} \) = 1.75 and \( \Delta t \) =10
MCS. The insets show the (rather sharp) variations of \( <E_{\mathrm{coop}}> \)
and \( <E^{2}_{\mathrm{coop}}> \) at the transition point, giving
rise to the dip there in \( \Delta E_{\mathrm{coop}} \). 

Fig. 5 (a) The variation of \( \textrm{d}m_{w}/\textrm{d} \)\( h_{p} \)
across two different points on the phase boundary: for \( T \) =1.50
and \( \Delta t \) = 2 MCS (open symbol) and \( T \) = 1.00 and
\( \Delta t \) = 12 MCS (solid symbol); (b) The variation of \( \mathrm{d}E_{\mathrm{tot}} \)\( /\textrm{d} \)\( T \)
across two different points on the phase boundary: for \( h_{p} \)
= 1.25 and \( \Delta t \) =10 MCS (open symbol) and \( h_{p} \)
= 1.75 and \( \Delta t \) = 10 (solid symbol).
\newpage

\vspace{0.3cm}
{\centering \rotatebox{90}{\includegraphics{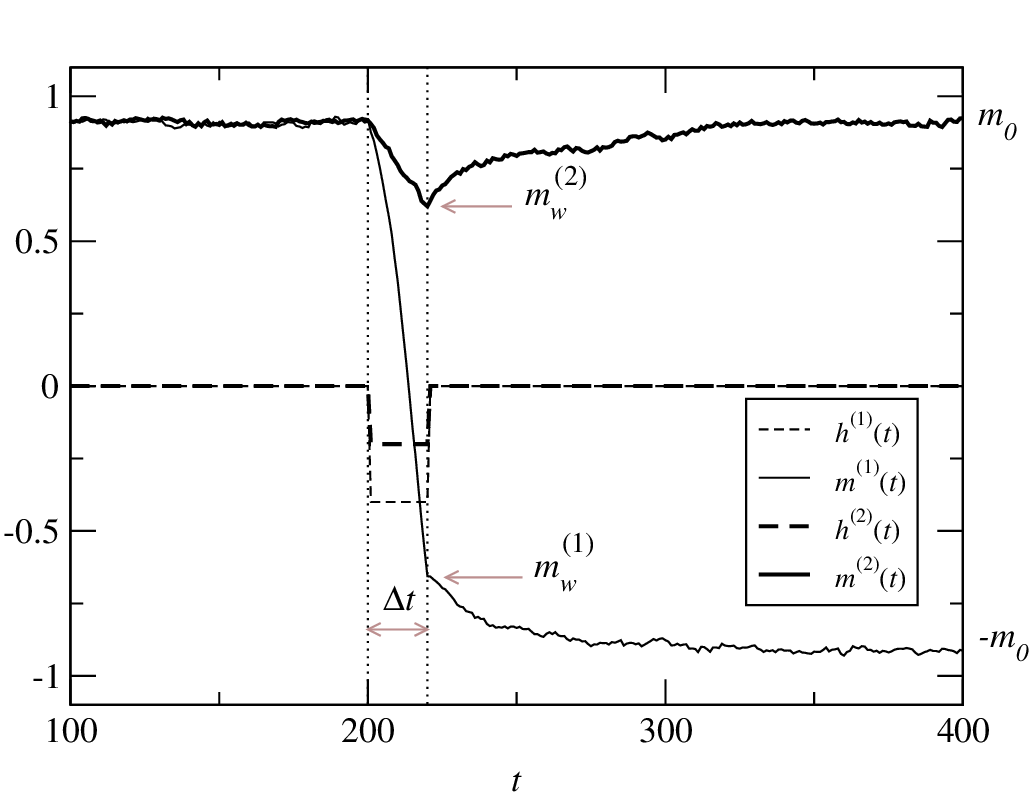}} \par}
\vspace{0.3cm}

\textbf{\LARGE Fig. 1 Chakrabarti et al}{\LARGE \par}
\newpage

\vspace{0.3cm}
{\centering \rotatebox{90}{\includegraphics{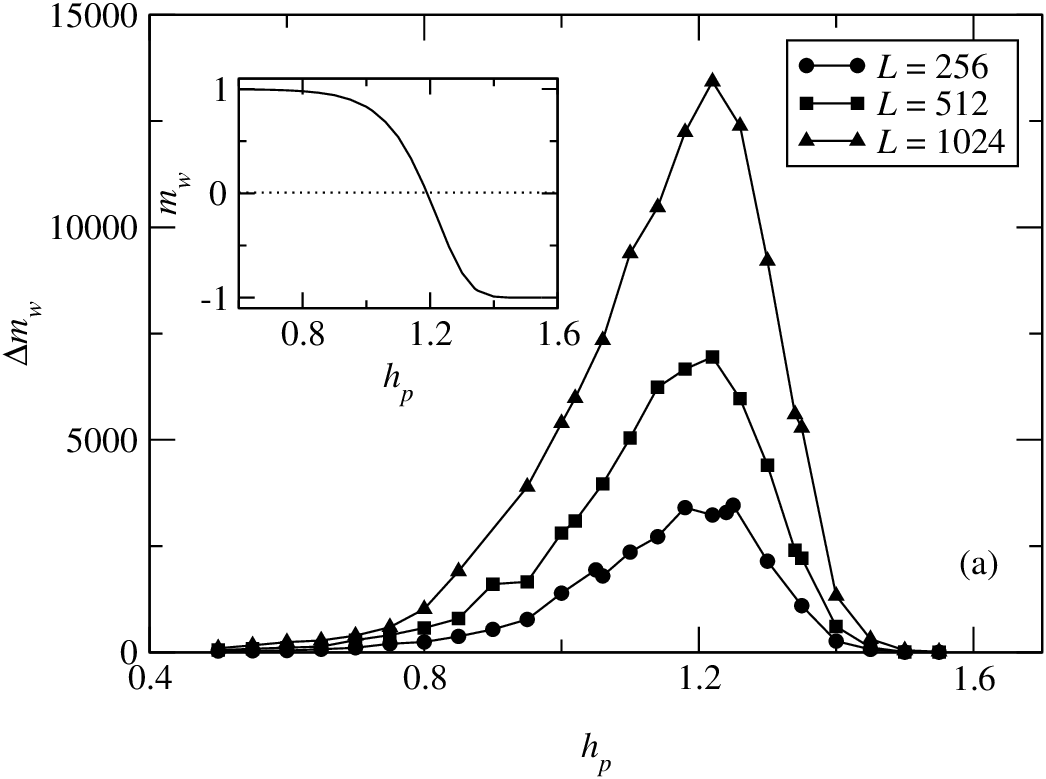}} \par}
\vspace{0.3cm}

\textbf{\LARGE Fig. 2 Chakrabarti et al}{\LARGE \par}
\newpage

\vspace{0.3cm}
{\centering \rotatebox{90}{\includegraphics{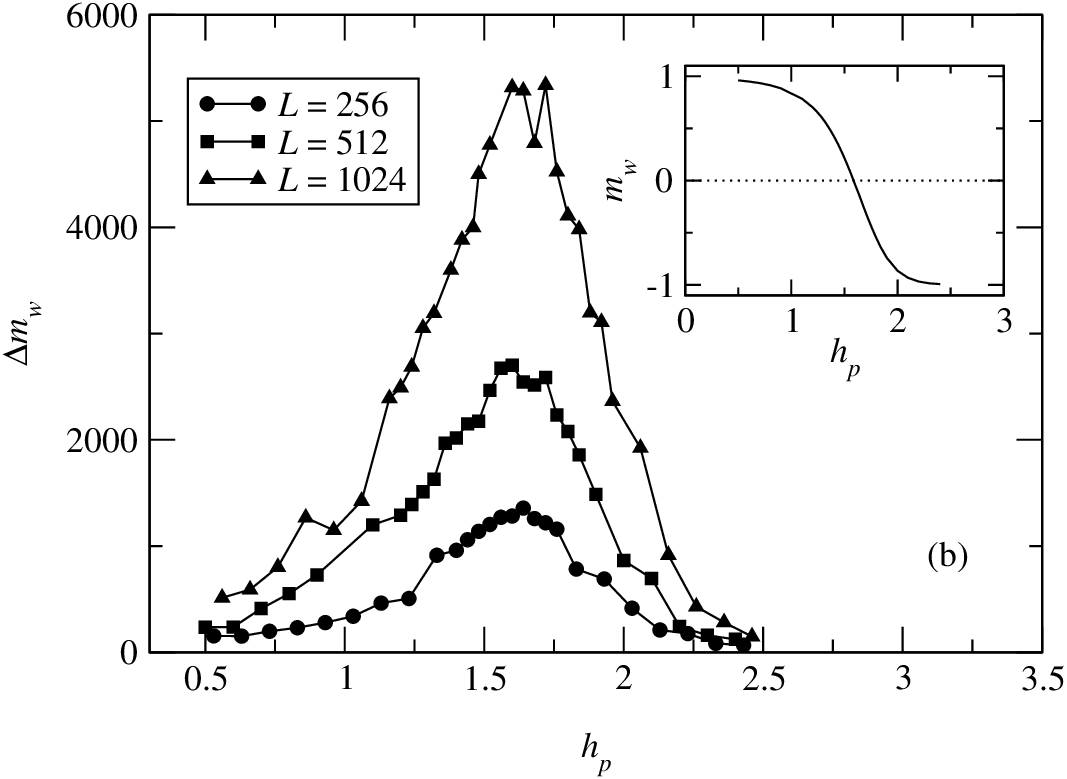}} \par}
\vspace{0.3cm}

\textbf{\LARGE Fig. 2 Chakrabarti et al}{\LARGE \par}
\newpage

\vspace{0.3cm}
{\centering \rotatebox{90}{\includegraphics{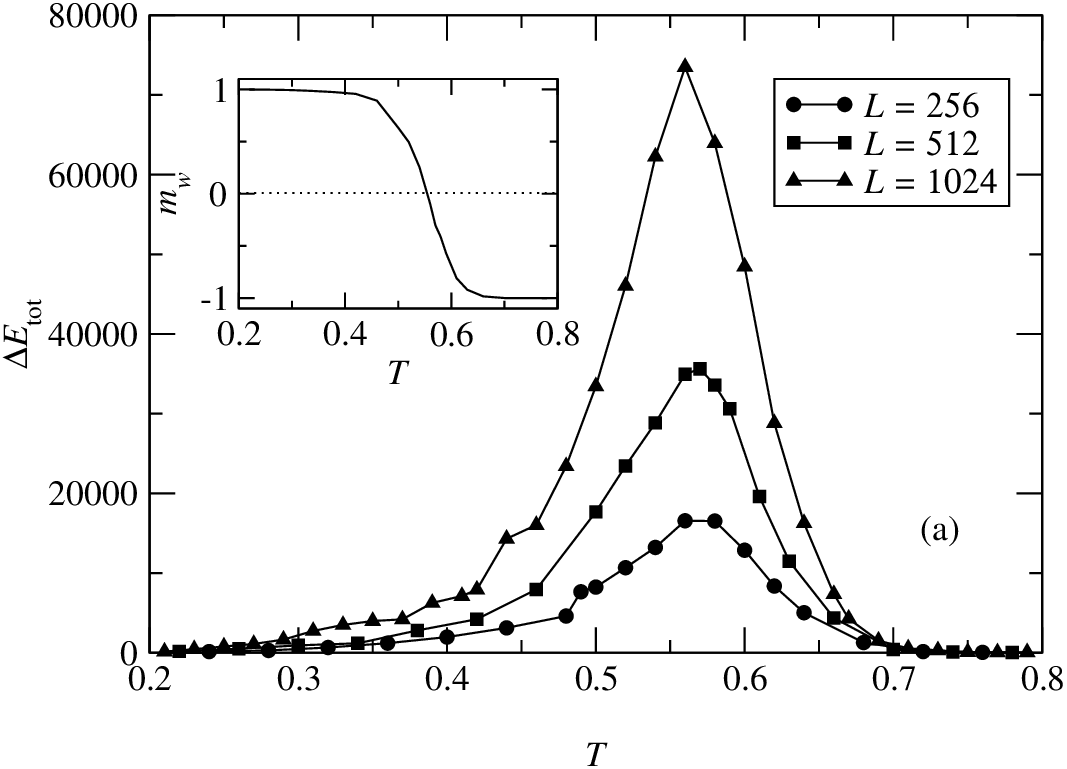}} \par}
\vspace{0.3cm}

\textbf{\LARGE Fig. 3 Chakrabarti et al}{\LARGE \par}
\newpage

\vspace{0.3cm}
{\centering \rotatebox{90}{\includegraphics{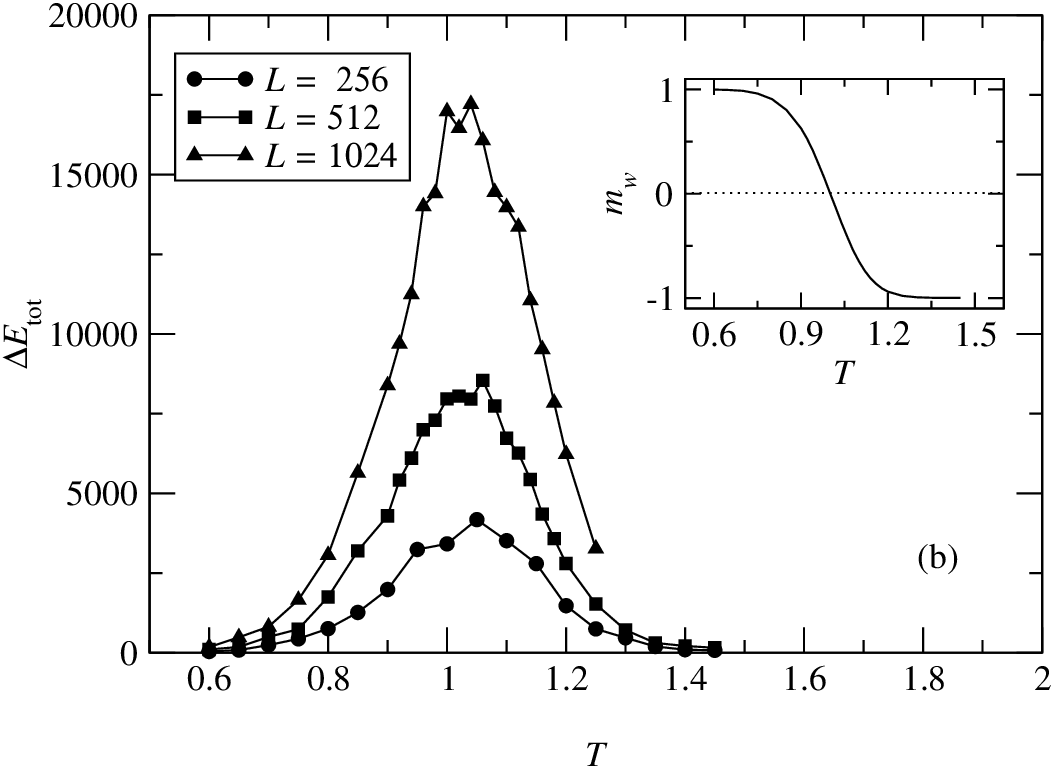}} \par}
\vspace{0.3cm}

\textbf{\LARGE Fig. 3 Chakrabarti et al}{\LARGE \par}
\newpage

\vspace{0.3cm}
{\centering \rotatebox{90}{\includegraphics{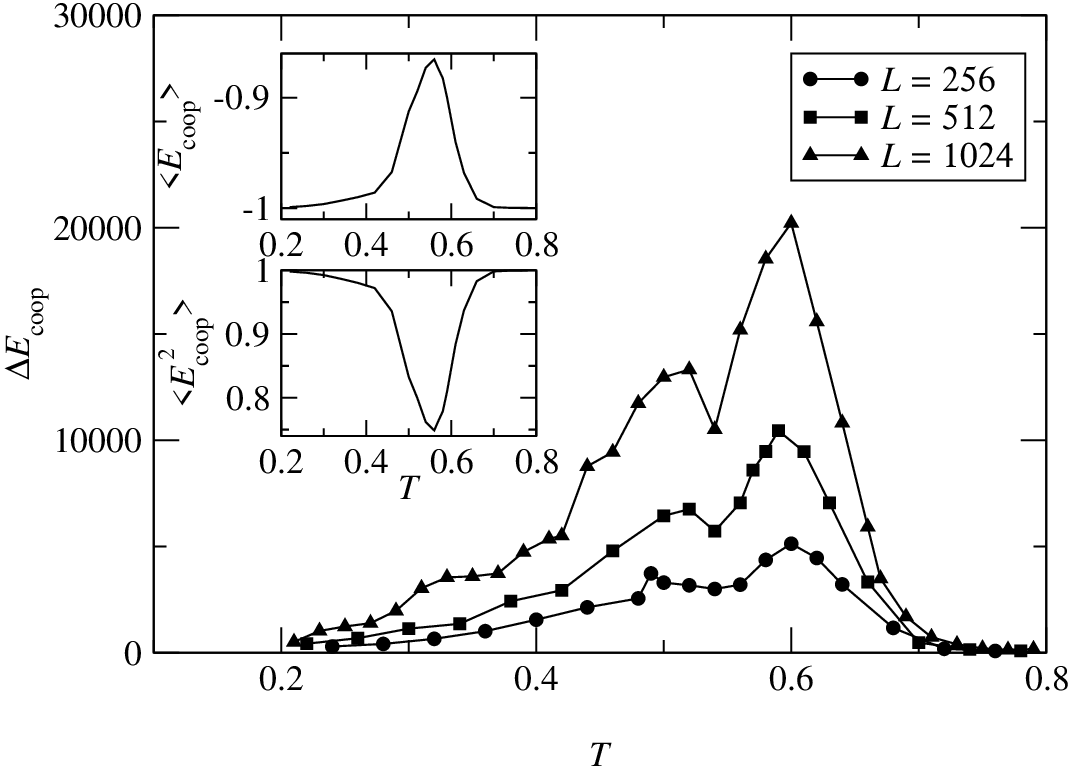}} \par}
\vspace{0.3cm}

\textbf{\LARGE Fig. 4 Chakrabarti et al}{\LARGE \par}
\newpage

\vspace{0.3cm}
{\centering \rotatebox{90}{\includegraphics{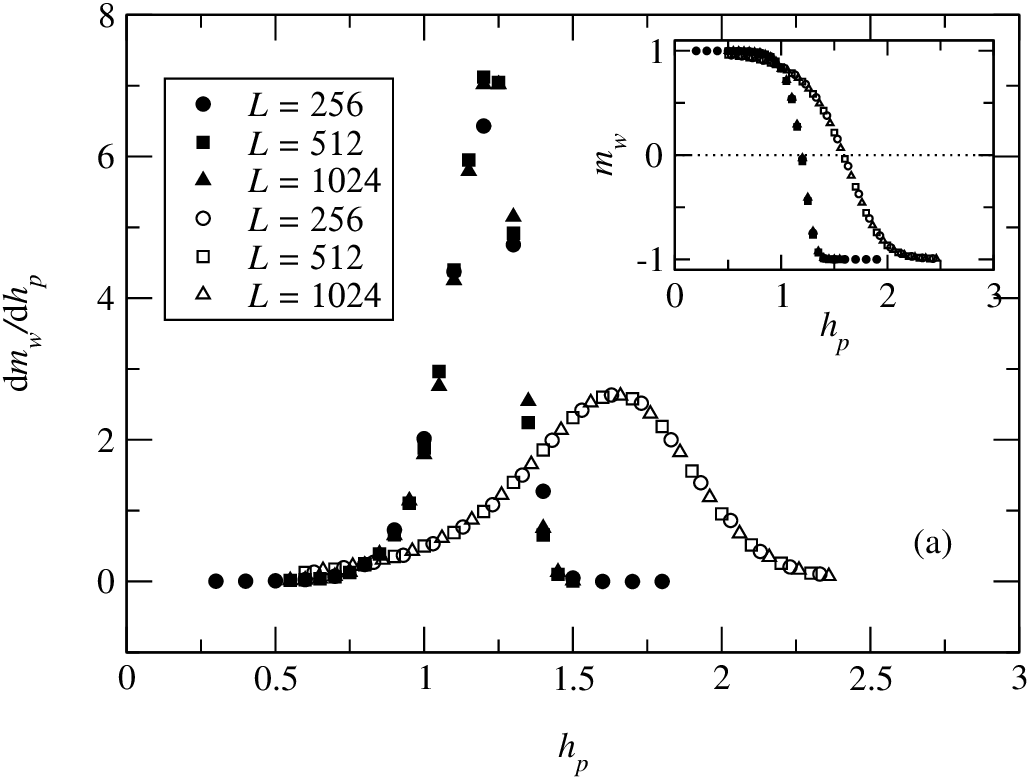}} \par}
\vspace{0.3cm}

\textbf{\LARGE Fig. 5 Chakrabarti et al}{\LARGE \par}
\newpage

\vspace{0.3cm}
{\centering \rotatebox{90}{\includegraphics{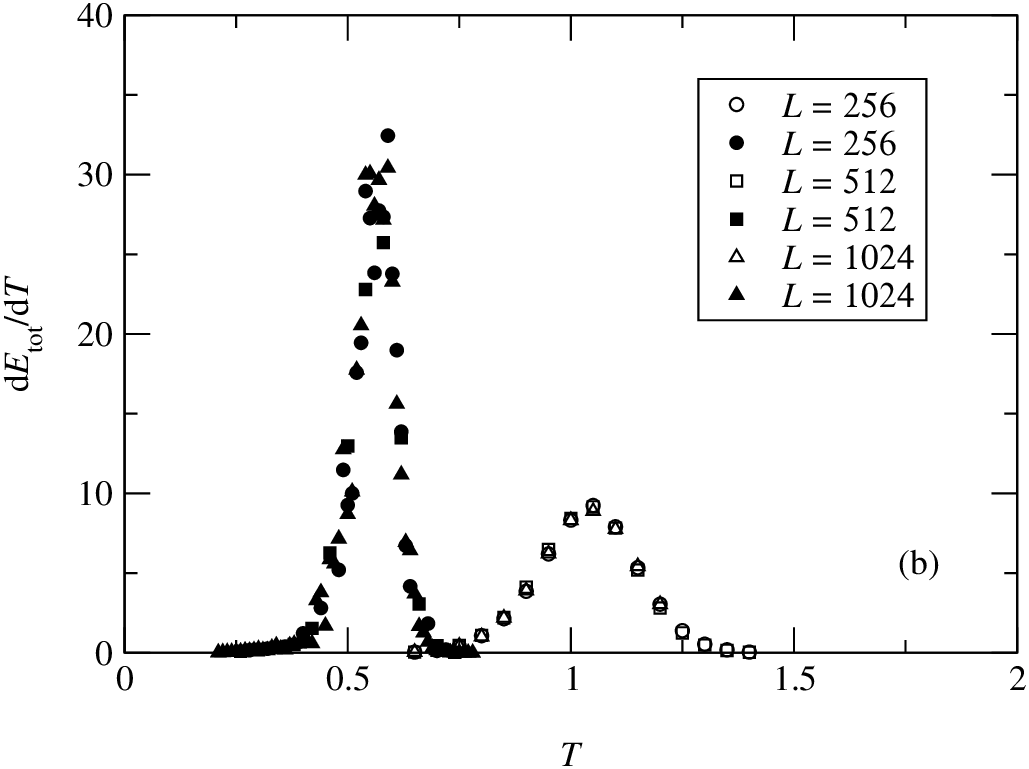}} \par}
\vspace{0.3cm}

\textbf{\LARGE Fig. 5 Chakrabarti et al}
\end{document}